# Fiber lasers: a power-scalable coherent light source for applications in space

Oliver de Vries
*Fraunhofer Institute for Applied Optics and Precision Engineering*
Albert-Einstein-Str. 7, 07745 Jena, Germany
Oliver.deVries@iof.fraunhofer.de

Marco Plötner
*Fraunhofer Institute for Applied Optics and Precision Engineering*
Albert-Einstein-Str. 7, 07745 Jena, Germany
Marco.Ploetner@iof.fraunhofer.de

Thomas Schreiber
*Fraunhofer Institute for Applied Optics and Precision Engineering*
Albert-Einstein-Str. 7, 07745 Jena, Germany
Thomas.Schreiber@iof.fraunhofer.de

Ramona Eberhardt
*Fraunhofer Institute for Applied Optics and Precision Engineering*
Albert-Einstein-Str. 7, 07745 Jena, Germany
Ramona.Eberhardt@iof.fraunhofer.de

Andreas Tünnermann
*Fraunhofer Institute for Applied Optics and Precision Engineering*
Albert-Einstein-Str. 7, 07745 Jena, Germany
Andreas.Tuennermann@iof.fraunhofer.de

*Abstract*—Fiber lasers entered numerous applications due to their high efficiency and superior stability. Particularly in space, those lasers have to fulfill additional requirements set by the harsh environment, typically characterized by large temperature gradients, vibration and shock as well as different kinds of radiation. In this contribution, we will report on the development and realization of a fiber laser intended to operate as part of a Rendezvous and Docking 3D-Lidar system considering the aforementioned specialties. Power scaling of pulsed as well as continuous-wave fiber lasers is yet another way to exploit new application fields in space.

## I. INTRODUCTION

On August the 12th 2014 the Automated Transfer Vehicle "Georges Lemaître" (ATV-5), operated by the European Space Agency, was successfully docking to the International Space Station. Beside its main mission as resupply spaceship, an experimental payload called LIRIS (Laser Infra-Red Imaging Sensors) was given a ride with the objective of testing a future generation of navigation sensors technologies during the docking maneuver. One of these sensors was a 3D scanning Lidar provided by the German company Jena Optronik GmbH. Real 3D-images of non-cooperative objects can be processed using this time-of-flight measurement technology, see Fig. 1 [1].

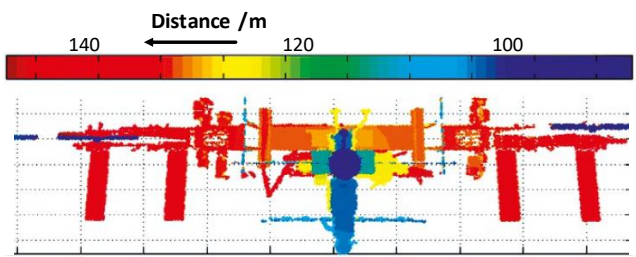

Fig. 1. Postprocessed 3D-Lidar image from data taken during the ATV-5 docking maneuver to the International Space Station (2014) [1].

To promote fast data acquisition, ns-pulses at a repetition frequency of ~40-kHz are provided by an Erbium-doped fiber laser (1550 nm), which was developed by the Fraunhofer Institute for Applied Optics and Precision Engineering (IOF) in Jena, Germany (Fig. 2). This fiber laser is capable to deliver up to 12 µJ of pulse energy and 4 kW of pulse peak power, but to meet eye-safety regulations during the LIRIS experiment, the average power was curbed to 10 mW (~30W pulse peak power).

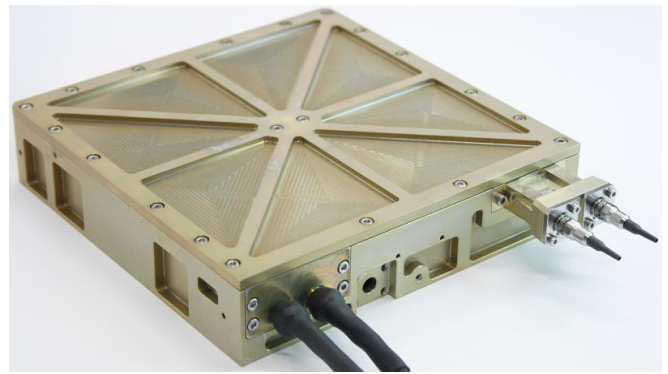

Fig. 2. Erbium-doped fiber laser as used in the LIRIS-experiment.

Fiber lasers start to emerge in space applications but are still considered as a new technology laden with teething troubles especially when it comes to radiation hardness and thus, long-term stability. In this contribution, we will show results of experimental and theoretical studies regarding the impact of radiation-induced absorption (RIA) on amplifier performance for different standard as well as radiation-hardened rare-earth doped amplifier fibers, which provide a practical guideline to optimize laser settings and ensure robust and reliable operation. Additionally, by incorporating a double-clad large mode area (LMA) fiber as main amplifier pulse energy can be scaled up to 100 µJ (>35 kW pulse peak power).

## II. PREAMPLIFIER STAGE

### A. Experiment: Optimization and fiber benchmarking

An important aspect for long term operation in space is the influence of ionizing radiation, which typically leads to strong absorptions especially in active fibers [2,3]. As a result, radiation-resistant fibers are recommended [4,5]. The fiber-based preamplifier is intended to have a high degree of amplifier performance (gain $>10^3$) and needs compliance towards existing (standard) fiber-optical components. Thus, making comparative experiments between standard and radiation-hardened gain fibers within an otherwise identical testbed is an adequate means to evaluate the performance of a space-qualified fiber amplifier, which inherently involves varying splice losses due to differences in core-size and core-NA. For this reason, five Erbium-doped single-clad fibers were tested in an all-fiber experimental setup (Fig. 3).

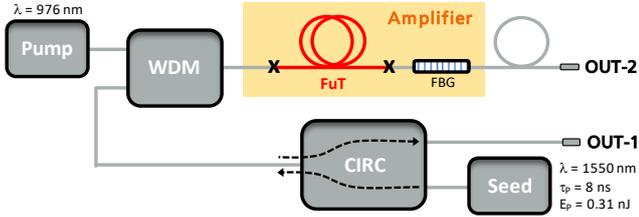

Fig. 3. Schematic of the all-fiber experimental setup used to acquire amplifier performance data for different gain fibers (FuT, fibers under test; FBG, fiber Bragg grating; WDM, wavelength-division-multiplexing; CIRC, optical circulator; OUT-1, main output; OUT-2, output for characterizing the residual emission).

Two of the five fibers have radiation-hardened properties whereas the three remaining standard-type fibers exhibit different parameters regarding core-size and doping concentration. Since the 1550-nm ns-pulsed distributed feedback laser seed diode is operated at a pulse repetition frequency of 20 kHz, the produced seed power is accordingly low (6.2 µW). This factor suggests that a small-core gain fiber is a better choice due to a lower saturation power. The 8-ns long pulses are electronically generated at the seed diode and propagate towards the double-pass amplifier passing through a fiber-optical circulator (CIRC) and wavelength-division-multiplexer (WDM). A 976-nm wavelength-stabilized single-mode laser diode pumps the amplifier fibers under test. To increase spectral purity, a fiber Bragg grating (FBG) with 3-nm bandwidth around the 1550 nm laser emission wavelength is utilized. The residual pump power is monitored and analyzed using output port OUT-2. Since this fiber amplifier stage is intended as an all-fiber solution, a high level of residual pump has to be avoided to prevent deterioration of the embedding material. The FBG back reflected pulse is amplified a second time and guided towards the output port OUT-1, where the main signal performance data can be acquired.

A comprehensive series of tests were performed for each fiber type by systematically varying the parameters of fiber length and pump power at a pulse repetition frequency of 20 kHz (average seed power 6.2 µW). From this data, the overall double-pass amplifier performance is derived in terms of amplification efficiency, signal-to-noise ratio (e.g. ASE-suppression) and (critical) unabsorbed pump power. The signal-to-noise ratio is calculated from the optical spectrum at OUT-1. The signal is defined as the spectral power within a 1-nm broad spectral window centered around the peak emission wavelength (1550 nm) whereas the noise comprises the remaining spectral power between 1500–1600 nm. Reason for a deterioration of this parameter is mainly an increasing ASE-level as well as the impact of the nonlinear effect of degenerated four-wave-mixing, which preferably occur in longer fibers at anomalous dispersion wavelengths and accordingly high peak powers. Due to its function as a reliable and robust first-stage high-gain pre-amplifier, we define the optimum performance as: (A) average output power >10 mW at OUT-1 (gain >1600, >32 dB), (B) residual pump power <5 mW at OUT-2 and (C) signal-to-noise ratio >97 % at OUT-1. The results reveal a noticeable change of the optimum operation range between each fiber type. One of the radiation hard fibers demonstrates a balanced performance with a broad operating optimum, which features moderate doping (equivalent to a moderate fiber length) and a small core-size (Fig. 4).

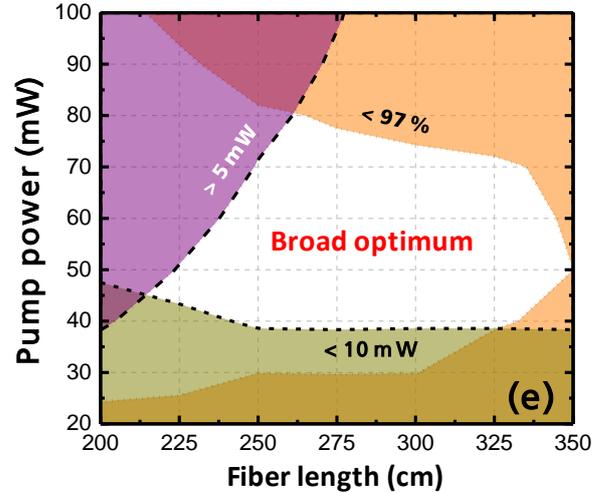

Fig. 4. The figure shows the summarized performance of the best benchmarked fiber. Within the white area, an optimum in terms of efficiency, signal-to-noise ratio and suppressed residual pump is reached (green area, average power <10 mW; violet area, residual pump power >5 mW; orange area, signal-to-noise ratio <97 %).

Two practical design rules are deduced from the results. Firstly, an optimum performance is reached at a 1530-nm peak core-absorption around 15 dB/m and, secondly, a smaller (4-µm) core-diameter is preferable.

*B. Simulation: Dose-dependent amplifier performance*

The exposure of (especially active) fibers to ionizing radiation will induce background losses, which in turn reduce the laser performance [6]. For this reason, the performance will alter during lifetime. To predict the radiation dose dependent gain reduction, a complete model based on the rate equation under consideration of amplified spontaneous emission (ASE) has been set up, able to simulate the influence of radiation-induced absorption of the fiber on the amplifier performance [7,8]. The numerical analysis of the system is carried out under steady-state approximations, thus, no temporal saturation effects are included. Since the model does not combine rate equation with the nonlinear Schrödinger equation, the effect of four-wave mixing is currently omitted. For this reason, no signal-to-noise ratio is analyzed in our theoretical investigation.

Own measurements on Er-doped fibers revealed a factor of 5.2 difference in radiation-induced absorption coefficient between 980-nm pump and 1550-nm signal wavelength. Based on this premise and a given background loss of 0.05 dB/krad at 1550 nm, the loss at pump wavelength will increase at a rate of 0.26 dB/krad accordingly. For the simulation, we chose again an average seed power of 6.2 µW. To verify a realistic representation of the numerical results, model parameters were chosen to fit the measurements of Fig. 4 at best. Figure 5 shows the output power as a function of fiber length and total radiation dose. At approximately 20 krad, the maximum output power has dropped by 10 dB. In the process, the position of the maximum is shifting to a shorter fiber length too, e.g. from ~3.5 m (no radiation) to ~1.75 m (20 krad).

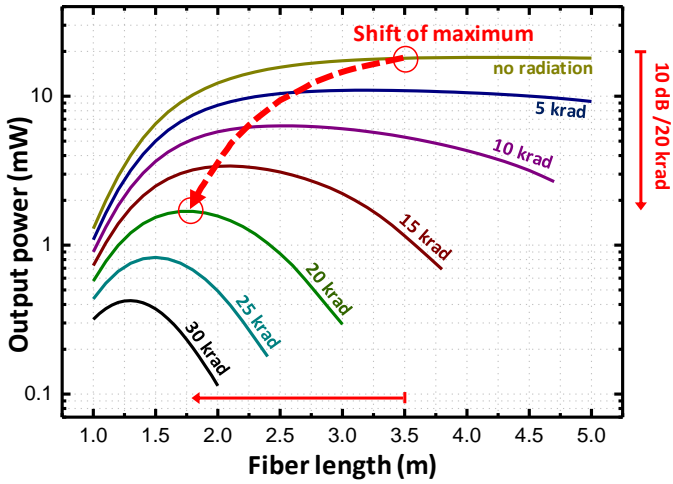

Fig. 5. Output power at OUT-1 for different radiation doses vs. fiber length at constant pump power of 60 mW. The optimum is shifting to shorter fiber lengths when higher doses are accumulated.

As plotted in Fig. 6, the radiation-induced gain loss can partly be compensated with a higher pump rate. For instance, the degradation from 10 to 20 krad of a 1.75 m long amplifier fiber can be counterbalanced by increasing the pump power from 100 to 200 mW.

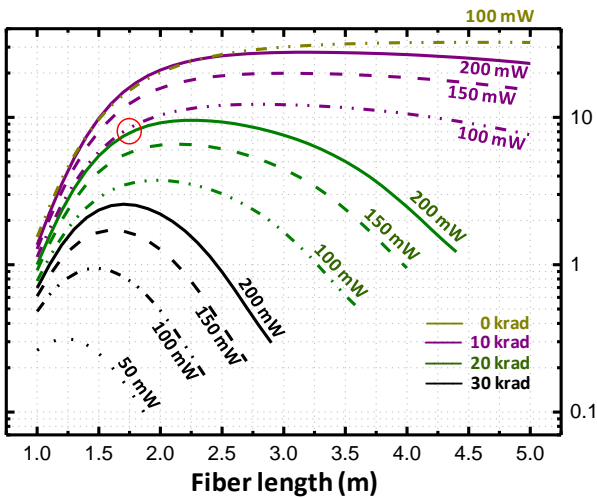

Fig. 6. Output power at OUT-1 for different radiation doses vs. fiber length at different pump powers which can partly compensate for radiation-induced gain losses (example marked).

The result suggests that for high expected radiation levels short fiber lengths and higher pump powers are most favorable. A convenient means is the implementation of a pump power control, which can partly compensate for radiation-induced gain reduction and enabling the use of shorter fibers by reducing the pump power at the beginning and gradually adjusting the amplifier performance during mission. Furthermore, the simulated powers can be used as seed input to estimate the performance of subsequent (fiber-) amplifier stages under the same radiation exposure. It should be noted, that the simulation does not include the implication of shielding and the compensating effect of photobleaching of radiation-induced color centers by 980-nm pumping [9] and hence can be considered as a worst-case scenario. Taking these issues into account, even higher radiation doses can be compensated by providing an adequate housing and/or allowing for photobleaching measures during flight.

For more details on the described preamplifier experiments and simulations, see [10].

### III. MAIN AMPLIFIER STAGE: SCALING PULSE ENERGY

The current space-grade fiber main amplifier stage enables the generation of 12 µJ pulses by using standard telecom fiber-technology. Limiting factors at this performance level are somewhat indistinct since the laser is intended to operate under safe conditions but basically two effects hinder further scaling efforts, energy saturation and the non-linear effect of four-wave-mixing, respectively. The latter will produce signal and idler photons outside the center wavelength and thus, reduce the signal-to-noise ratio while the former leads to a shortening of pulse duration. A small fiber core and, compared to e.g. crystal-based rod lasers, a relatively low doping concentration limit the total number of rare earth doped ions and hence the extractable energy. As a result, the inversion (gain) will change over pulse duration, with a higher gain at the leading and a lower gain at the trailing edge of the pulse. Shorter pulses in turn will generate higher peak powers and will pronounce the repercussion of four-wave-mixing again. A possible solution provides the utilization of large-mode-area (LMA) double-clad fibers (DCF) [11–13]. By increasing the core size some benefits can be achieved. Compared to standard DCF, the larger core of a LMA-DCF reduces the power density and the pump-absorption length of the fiber, both resulting in less light-matter interaction and a reduction of nonlinearities. Also the volume of actively doped glass increases and concomitantly the capability to store energy.

In our experiments, we use a 30-µm core Erbium-Ytterbium co-doped fiber with a 300-µm pump core. To match the mode field diameter of the preamplifier to the LMA-DCF, a mode-field adapter was produced using in-house $CO_2$ glass processing technology [14,15]. This technology also provides the future perspective to implement LMA-DCF in a rugged, alignment-free all-fiber monolithic setup. At about 12 W of 980-nm pump power a pulse energy of 100 µJ can be extracted with pulse repetition frequencies between 20–40 kHz with a beam quality factor of $M^2=1.25$ (Fig. 7). Pulse duration retains at 3.3 ns up to 50 µJ and starts to get shorter with rising energy down to 2.7 ns at 100 µJ indicating the onset of gain saturation. This pulse shortening leads to a nonlinear slope of the corresponding pulse peak power, shown in Fig. 8.

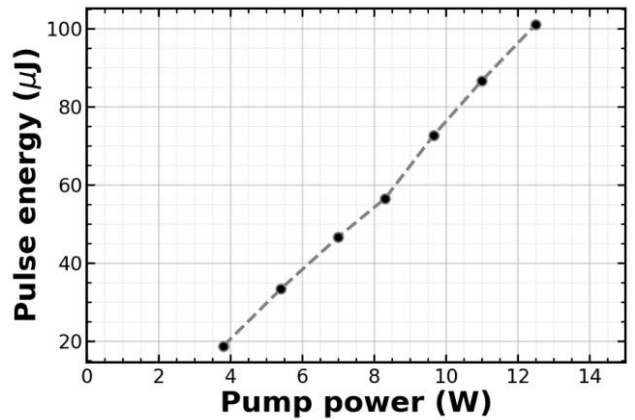

Fig. 7. Pulse energy as function of total pump power (PRF=20 kHz). The light is 1-nm band pass filtered around the center-wavelength of 1550-nm.

The experiment shows that by utilizing LMA-DCF a scaling of pulse energy and pulse peak power by one order of magnitude is possible. Since average power is scalable, 100 µJ are expected even at significantly higher repetition frequencies (>100 kHz). In future steps, a signal feed-through pump combiner adapted to this particular LMA-DCF would allow for the realization of an all-fiber setup, ideally suited for applications in aerospace.

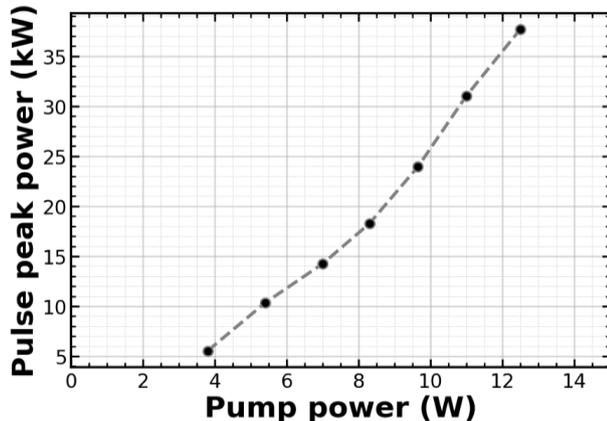

Fig. 8. Pulse peak power as function of total pump power (PRF=20 kHz). This plot takes into account the pulse shortening due to energy saturation effects.

IV. CONCLUSION

In this contribution, we presented the past, current and future activities regarding the space-suitable fiber laser development at Fraunhofer IOF. By applying LMA-DCF as booster amplifier, 100-µJ/35-kW ns-pulsed fiber lasers can be realized in a potentially monolithic and hence alignment-free packaging. Beside Lidar-type applications, the scaling of fiber lasers go along with ambitious ideas like long-distance broadband laser communication, power transmission to and from space [16] as well as visionary endeavors like the space-debris removal via ICAN-architecture [17].

ACKNOWLEDGMENT

We thank our partners at Jena Optronik GmbH for their long-lasting cooperation. The work presented was partly performed within the respective project by DLR German Space Agency under project no. 50 RA 1511 with funds from the German Ministry for Economy and Technology (BMBF) by enactment of the German Bundestag.